\documentclass[prl,twocolumn]{revtex4-2}

\usepackage{graphicx}
\usepackage{multirow}
\usepackage{dcolumn}
\usepackage{bm}
\usepackage[normalem]{ulem}
\usepackage{amsmath}
\usepackage{amsfonts}
\usepackage{amssymb}
\usepackage{color}
\usepackage{colordvi}
\usepackage{dcolumn}

\newcommand{\sgn}{\mathrm{sgn}}

\begin{document}

\title{Probing the $\nu=2/3$ fractional quantum Hall edge by momentum-resolved tunneling}
\date{\today }
\author{Hendrik Meier$^{1}$, Yuval Gefen$^{2}$, and Leonid I. Glazman$^{1}$}
\affiliation{
$^1$Department of Physics, Yale University, New Haven, Connecticut 06520, USA\\
$^2$Department of Condensed Matter Physics, Weizmann Institute of Science, Rehovot 76100, Israel
}

\begin{abstract}
The nature of the fractional quantum Hall state with filling factor~$\nu=2/3$ and its edge modes continues to remain
an open problem in low-dimensional condensed matter physics. Here, we suggest an experimental setting to probe the
$\nu=2/3$ edge by tunnel-coupling it to a $\nu=1$ integer quantum Hall edge in another layer of a two-dimensional electron gas (2DEG).
In this double-layer geometry, the momentum of tunneling electrons may be boosted by an auxiliary magnetic field parallel to the two planes of 2DEGs. 
The threshold behavior of the current as a function of bias voltage and the boosting magnetic field yields information about 
the spectral function of the $\nu=2/3$ edge, and in particular about the nature of the chiral edge modes. Our theoretical analysis 
accounts also for the effects of Coulomb interaction and disorder.
\end{abstract}

\maketitle
\paragraph*{Introduction.} 
In the conditions of quantum Hall effect, 
the compressible regions remain only 
at the edges of a sample
and are known as edge states \cite{halperin_81,milliken}. 
Edges of \emph{fractional} Hall states are strongly-correlated chiral electron liquids \cite{wen_review}. The situation becomes even more intriguing for systems
with filling factors different from those of Laughlin states, for which $\nu= 1/m$ with odd integer $m>0$. Here we consider
the particular case of a quantum Hall system with filling factor~$\nu=2/3$.

A prominent conjecture \cite{macdonald,fradkin_book} considers
the $\nu=2/3$ edge \cite{haldane_83,halperin_84} as composed of two spatially separated edge channels:
an outer channel with filling factor~$\nu=1$ and an inner counter-propagating one corresponding to
a $\nu=1/3$ liquid of hole states.
Coulomb interactions between the two channels and backscattering off disorder
enrich the physical picture and in the low-energy limit drive the system into a state with universal two-terminal and Hall conductance~$G=2e^2/3h$ \cite{kfp}.
Theory \cite{kfp} predicts two effective edge modes, a charge-carrying mode and a counter-propagating
neutral one. Recent shot noise measurements at a quantum point contact \cite{bid_mahalu_2010} have provided indirect evidence for the 
existence of such a neutral mode and a mechanism of upstream heating by neutral currents \cite{bid_mahalu_2010,takei_rosenow,yacoby_2012}.
However, two observations put the picture of two counter-propagating modes in question and favor the possibility of 
two co-propagating $\nu=1/3$ modes as alternatively suggested long ago \cite{beenakker}. One is the observation of a $G=e^2/3h$ plateau 
\cite{changcunningham,bid_mahalu_2009} in the conductance through quantum point contacts. 
The second is the effective charge, detected through shot noise measurements \cite{bid_mahalu_2009}, which crosses over from $e/3$
at higher temperature to $2e/3$ at lower ones. A ``unified'' theory
has recently \cite{wangmeirgefen} been proposed in terms of a four-channel model for a reconstructed edge.
In this situation of competing theories, direct experimental evidence about the internal structure of the
$\nu=2/3$ edge is called for.

In this paper, we are suggesting a bilayer experiment to investigate the spectral function and, in particular,
the nature of the chiral modes of the $\nu=2/3$ edge. In this experiment, the fractional quantum Hall edge is probed
by momentum-resolved tunneling into or from the edge of an integer quantum Hall state with filling factor~$\nu=1$, which we understand rather well.
Applying an \emph{in-plane} magnetic field~$B_y$ allows one to extract the spectral function upon
measuring the current~$I$ as a function of~$B_y$ and bias voltage~$V$ in a two-terminal setting. We show how
the geometry of the edge channels corresponds in the $V$--$B_y$ plane to a pattern of equidistant \emph{valleys} of current~$I$, where lines of non-analyticity
with exponents characteristic to the filling factor intersect.

The suggested experiment on the $\nu=2/3$ edge is inspired by experiments on tunnel-coupled parallel quantum wires \cite{auslaender_etal}.
In the latter setup, measuring the tunnel current as a function of bias voltage and a transverse magnetic field provided 
direct information about the threshold lines (in energy--momentum space) for the spectral function
and, hence, about the velocities of the spin and charge
modes of the one-dimensional electron liquid \cite{giamarchi}. In the context of the quantum Hall effect, similar
settings using momentum-conserved tunneling have been discussed in Refs.~\cite{zuelicke,yang_1,feldman,yang_2}. 
Bilayer settings like the one we are suggesting are to be distinguished from
tunnel-coupled lateral quantum Hall systems \cite{kang}.

\begin{figure}[t]
\centerline{\includegraphics[width=0.9\linewidth]{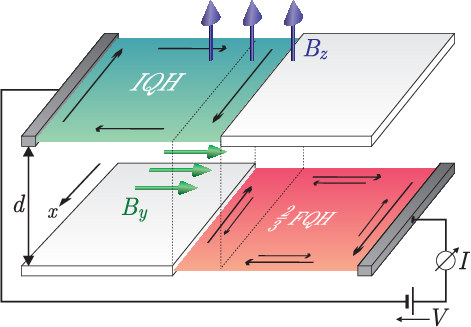}}
\caption{(Color online) Suggested experimental setup to study the~$\nu=2/3$ fractional quantum Hall (FQH) edge. The tunnel current~$I$ between the FQH edge and the probing $\nu=1$ integer quantum Hall (IQH) edge is measured as a function of bias voltage~$V$ and transverse magnetic field~$B_y$, which boosts the momentum of tunneling electrons
by $Q=eB_y d$.}
\label{fig01}
\end{figure}

\paragraph*{Setting.} Figure~\ref{fig01} presents the double-layer layout we suggest for probing the $\nu=2/3$ state. Each layer contains a 2DEG with suitable individual 
gating and doping such that the magnetic
field~$B_z$ establishes a $\nu=2/3$ state in the lower layer and a $\nu=1$ state in the upper one.
The distance~$d$ between the layers is chosen such that the layers are coupled via electron
tunneling. The specific feature of our setting is that the edges of the two quantum Hall states are aligned on top of each other.
This may be achieved by applying voltages to a top-right gate and a bottom-left gate that deplete the right part of the upper and the left part of the lower 2DEG.
Full depletion of one layer's halfplane without affecting the other layer is possible for screening lengths of the order of the Bohr radius,
which in realistic experiments can be the case \cite{eisenstein_1}. 
In the absence of quantum Hall features ($B_z=0$), momentum-resolved tunneling between two parallel 2DEGs was experimentally realized \cite{eisenstein_2} for 
$d\approx 70\ \mathrm{\AA}$ and in-plane magnetic fields up to~$B_y\sim 8\ \mathrm{T}$.
Scanning over a width of the magnetic length~$\ell_{B_z}=\sqrt{\hbar/eB_z}\sim 100\ \mathrm{\AA}$ requires fields~$B_y$ up to $10\ \mathrm{T}$.
In order to avoid such strong transverse fields, one may equivalently 
adjust the gate voltages to move the edges by~$\Delta y$ \cite{wei}, corresponding to a momentum boost~$\hbar\Delta y/\ell_{B_z}^2$,
and resort to a~$B_y$ of
smaller magnitude for fine-tuned momentum scans close to a valley.

\paragraph*{Model.} The upper integer quantum Hall edge,
which serves to probe the $\nu=2/3$ edge, is described using a simple model of chiral electrons with spectrum~$\varepsilon_{0,k}=u_0k-\varepsilon_0$ 
that propagate in a positive (upward) $x$-direction.
For the $\nu=2/3$ edge, we adopt the low-energy fixed-point theory by Kane, Fisher, and Polchinski (KFP) \cite{kfp} for 
a theoretical discussion of the suggested experiment.
KFP assume in the bare picture
an exterior downward-propagating $\nu=1$ and an inner upward-propagating $\nu=1/3$ edge channel \cite{wen_review,macdonald,fradkin_book} associated with bosonic fields
$\phi_1(x)$ and $\phi_2(x)$, respectively. These satisfy the commutation relations $[\phi_i(x),\phi_j(x')]=(\mathrm{i}\pi\delta_{ij}/K_i)\ \sgn(x-x')$ with
$K_1=-1$ and $K_2=3$.

\begin{figure}[t]
\centerline{\includegraphics[width=0.9\linewidth]{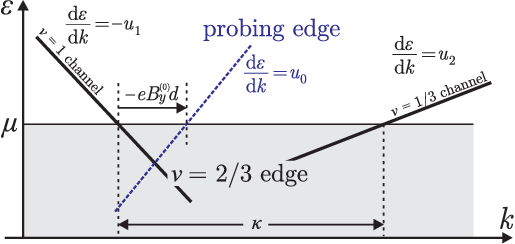}}
\caption{(Color online) Spectrum of the $\nu=2/3$ edge channels (solid lines) containing a branch for the outer downward-propagating $\nu=1$ channel and 
the the inner counter-propagating $\nu=1/3$ channel. The dashed blue line represents the edge of the probing $\nu=1$ integer quantum Hall edge of the upper layer.}
\label{fig02a}
\end{figure}

The clean $\nu=2/3$ edge is described by the Hamiltonian
\begin{align}
\hat{H}_{2/3} &= 
\int\frac{\mathrm{d}x}{4\pi}\ 
\big[
u_1(\nabla\phi_1)^2 + 3u_2(\nabla\phi_2)^2
+ 2u_{12}\nabla\phi_1\nabla\phi_2
\big]\ .
\label{eqn11}
\end{align}
We assume that velocities~$u_1$ and~$u_2$ have already been renormalized by intrachannel Coulomb interactions.
If interchannel Coulomb interactions are absent ($u_{12}=0$), the Hamiltonians for
the probing edge and the $\nu=2/3$ edge lead to the spectrum in Fig.~\ref{fig02a}. The 
spatial separation of the two channels inside the $\nu=2/3$ edge implies a (gauge-invariant) distance~$\kappa$ between
their Fermi points in canonical momentum space. It is determined by the chemical potential~$\mu$, the 
field~$B_z$ establishing the quantum Hall regime, and details of the edge potential.

At finite~$u_{12}$, the independently propagating modes are given, instead of~$\phi_{1,2}$, 
by~$\phi_+=\sqrt{\eta-1}\phi_1+\sqrt{3\eta}\phi_2$ and $\phi_-=\sqrt{\eta}\phi_1+\sqrt{3(\eta-1)}\phi_2$,
propagating with velocities~$u_{\pm} = (u_2-u_1)/2 \pm (u_1+u_2)\gamma/2$. Herein, $\eta=(1+\gamma)/2\gamma$, $\gamma=\sqrt{1-c^2}$,
and $c=(2/\sqrt{3})\ u_{12}/(u_1+u_2)$. 
The coefficient~$\eta$ takes values 
between $\eta=1$ for~$u_{12}=0$ and $\eta=3/2$ at the low-energy fixed point \cite{kfp}. In terms of
the effective modes,
\begin{align}
\hat{H}_{2/3} &= 
\int\frac{\mathrm{d}x}{4\pi}\ 
\big[
-u_{-}(\nabla\phi_{-})^2 + u_{+}(\nabla\phi_{+})^2
\big]\ .
\label{eqn12}
\end{align}
We assume $u_+<u_0$, consistent with slow neutral modes.

At the low-energy fixed point $\eta=3/2$, charge transport only involves~$\phi_-$ whereas $\phi_+$ is a counter-propagating neutral mode \cite{kfp}. Reaching this fixed point requires an equilibration mechanism such
as backscattering off impurities. The most relevant Hamiltonian for interchannel backscattering reads 
\begin{align}
\hat{H}_D &= 
\int\mathrm{d}x\ 
\xi(x)\ 
\exp\big\{
-\mathrm{i}(\phi_1+3\phi_2)
\big\} + \mathrm{H.c.}
\label{eqn13}
\end{align}
Herein, the operator $\exp(-\mathrm{i}\phi_1)$ creates a quasi-particle of charge~$-e$ in the outer channel while $\exp(-3\mathrm{i}\phi_2)$
annihilates three quasi-particles, each of charge~$-e/3$, in the inner channel.
Following~KFP, we assume Gaussian disorder, $\langle\xi(x)\xi^*(x')\rangle= w\delta(x-x')$.
At~$\eta=3/2$, $\hat{H}_{D}= \int\mathrm{d}x
\ \xi(x)\exp(-\mathrm{i}\sqrt{2}\phi_+) + \mathrm{H.c.}$, so only the neutral mode is affected.

Finally, we choose a proper model for interedge tunneling. 
We assume the barrier between the probing and fractional edge homogeneous, implying momentum-conserving tunneling at zero transverse field~$B_y$.
A finite~$B_y$
boosts the momentum of the tunneling electron by~$Q=eB_y d$. Assuming tunneling matrix elements~$t_0 \delta_{kk'}$ 
in the space of canonical momentum~$k$, we are led to the tunneling Hamiltonian
\begin{align}
\hat{H}_{T} &= t_0 \sum_k \hat{\psi}^\dagger_{2/3,k}\hat{\psi}^{\phantom{\dagger}}_{0,k+Q} + \mathrm{H.c.}
\label{eqn15}
\end{align}
Here $\hat{\psi}_{0}$ is the electron field operator for
the probing edge. The most general field operator for the
annihilation of charge~$-e$ in the~$\nu=2/3$ edge has the form
\begin{align}
\hat{\psi}_{2/3}(x)&=
\frac{\mathrm{e}^{\mathrm{i}\phi_1(x)}}{(2\pi\alpha)^{1/2}}
\sum_{n=-\infty}^\infty A_n\mathrm{e}^{\mathrm{i}n\kappa x} \mathrm{e}^{-\mathrm{i}n[\phi_1(x)+3\phi_2(x)]}
\label{eqn16}
\end{align}
with amplitudes~$A_n$ and ultraviolet length cutoff~$\alpha$. 
If the channels were uncoupled, all~$A_n$ would be zero except for $A_0$ and~$A_1$. 
These two correspond to annihilation of charge~$-e$ in the~$\nu=1$ and $\nu=1/3$ channels, respectively. 
Terms different from these two additionally transfer integer multiples of charge~$e$ 
between the two channels, thus generating dipole excitations. We assume $A_n\sim 1$ for all~$n$ as is the case
in a generic Luttinger liquid \cite{giamarchi}, where~$\kappa$ is given by twice the Fermi momentum.

\paragraph*{Kinematic picture.} Before delving into effects of disorder and interaction, let
us discuss the kinematics of the setup in Fig.~\ref{fig01}.
A finite bias voltage~$V$ effectively shifts both the spectrum and the chemical potential of the probing edge (Fig.~\ref{fig02a}) in vertical direction
while a momentum boost by~$B_y$ corresponds to a horizontal shift of
the spectrum of the probing edge. There exists a unique value~$B_y^{(0)}$ such 
that interlayer tunneling couples the Fermi momentum state of the probing edge 
to that of the $\nu=1$ channel in the fractional edge. Henceforth,
we denote by~$B_y$ the magnetic field measured from~$B_y^{(0)}$. In $\hat{H}_{T}$, Eq.~(\ref{eqn15}), this
has already been assumed.

\begin{figure}[t]
\centerline{\includegraphics[width=\linewidth]{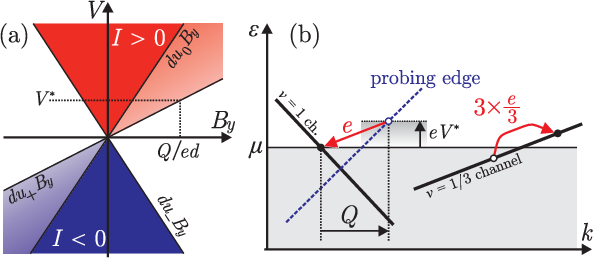}}
\caption{(Color online) Kinematic picture for tunneling between the probing edge and the $\nu=1$ channel of the $\nu=2/3$ edge. 
(a) Regions of non-zero tunnel current in the~$B_y$--$V$ plane.
(b) Interaction-induced tunneling at (positive) bias voltage~$V^*=Qu_+/e$.}
\label{fig02b}
\end{figure}

Neglecting interchannel Coulomb interactions, $u_-=-u_1$ and $u_+=u_2$.
For small $Q=eB_y d$, the probing edge is close to the outer channel of the $\nu=2/3$ edge,
and we need to retain only the term~$n=0$ in the sum of Eq.~(\ref{eqn16}).
Momentum-resolved tunneling between the edges requires the intercept
point~$k_\times$ of the edge bands to correspond to an occupied state in the shifted probing edge ($\varepsilon_{0,Q+k_\times}<\mu+eV$)
and to an empty one in the fractional quantum Hall edge ($\varepsilon_{1,k_\times}>\mu$) or vice versa. 
At zero temperature, this condition is met for pairs $(B_y,V)$ in the dark-colored regions in Fig.~\ref{fig02b}(a)
with boundaries defined by velocities~$u_-$ and~$u_0$. In these regions, Fermi's golden rule yields a current per unit length of value
$I=\pm et_0^2/(u_0+u_1)$ (see Appendix A).

Consider now the bias voltage~$V^*=Qu_+/e$ [Fig.~\ref{fig02b}(b)]. Clearly, $V^*$ is not strong enough for
momentum-resolved tunneling as discussed above, yet imagine an electron at the Fermi point of the probing edge tunneling 
into the lowest unoccupied state of the nearby fractional quantum Hall edge channel. This violates momentum and energy conservation. 
But, since the quotient of energy and momentum mismatches equals the slope~$u_+$ of the inner $\nu=1/3$ channel, 
an interaction-induced quasi-particle excitation in this inner channel can 
restore overall momentum and energy conservation. 
This Coulomb-supported tunneling works for voltages above $Qu_+/e$, annexing the light-colored zones in Fig.~\ref{fig02b}(a)
to the regions of non-zero current, whose boundaries are thus determined solely by the spectrum of the $\nu=2/3$ edge.

Similar considerations at $Q\sim\kappa$ lead to another \emph{valley} in the $B_y$--$V$ plane for tunneling between the probing edge and
the $\nu=1/3$ channel, corresponding to the $A_1$ term in Eq.~(\ref{eqn16}). Including all~$A_n$ leads to a pattern
of valleys situated at integer multiples of~$\kappa$. The additional valleys are analogs to the ``shadow
bands'' in one-dimensional systems \cite{penc}. We now turn to studying the current within the proper Luttinger-liquid formalism.

\begin{figure}[t]
\centerline{\includegraphics[width=\linewidth]{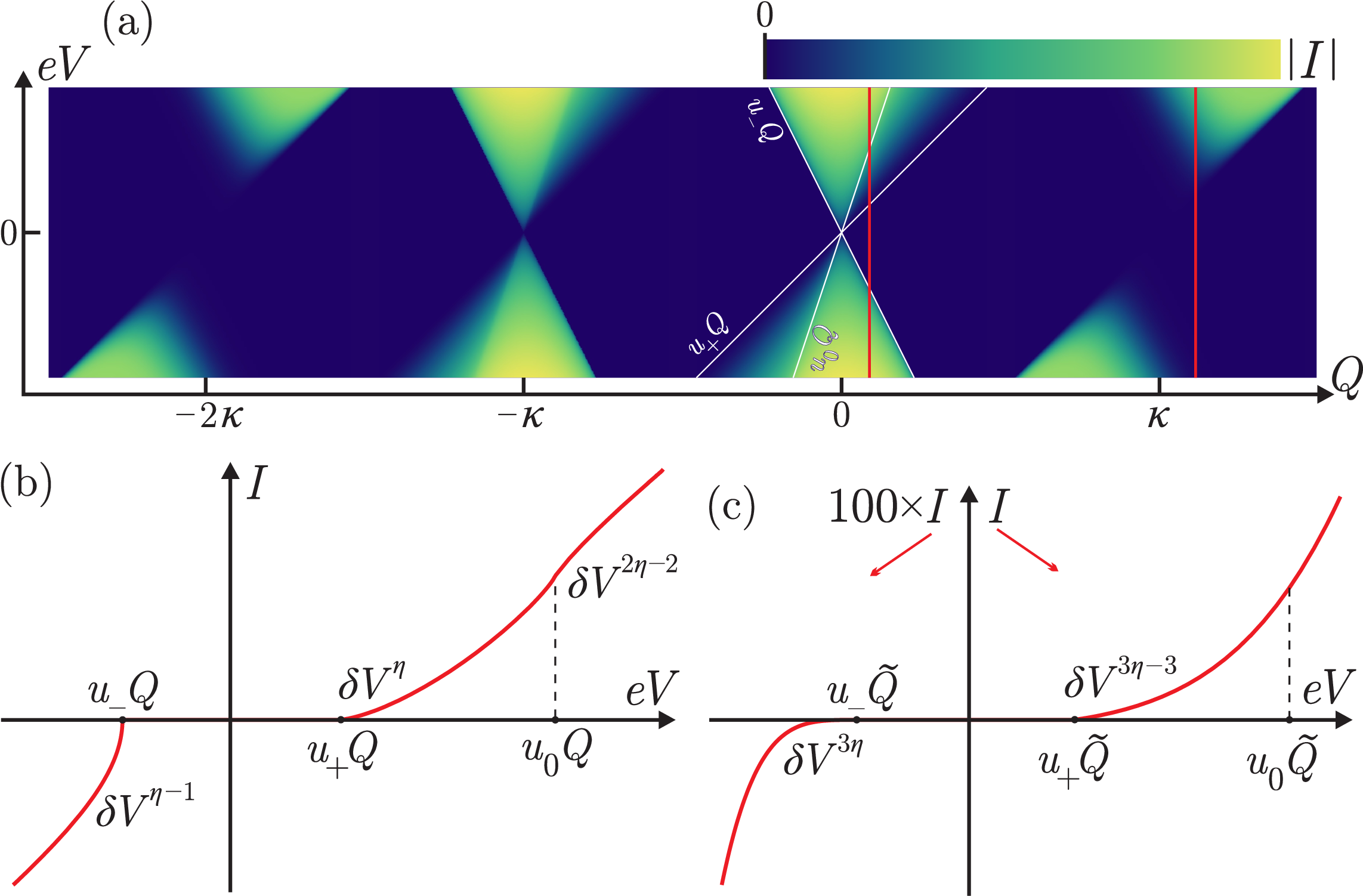}}
\caption{(Color online) (a) First four valleys for the current~$I(V,Q)$ where~$Q=eB_y d$. 
Coarse control of~$Q$ on the scale of~$\kappa$ is possible also by tuning the gate voltages.
(b), (c) $I(V)$ curves for valleys~$n=0$ and~$1$ and asymptotic power laws close to non-analytic lines, $V=u_\zeta Q/e +\delta V$ where $\zeta\in\{\pm,0\}$. At $\eta=3/2$, the power law~$\delta V^{2\eta-2}$ turns into $-\delta V\ln|\delta V|$.}
\label{fig03}
\end{figure}

\paragraph*{Current.} In linear-response theory with respect to interlayer electron tunneling~$\hat{H}_T$, we obtain the
tunnel current per unit length between the $\nu=2/3$ and the probing edge
by expanding $I=-2et_0 \sum_k\mathrm{Im}\langle\hat{\psi}^\dagger_{2/3,k}\hat{\psi}^{\phantom{\dagger}}_{0,k+Q}\rangle$ to order~$t_0^2$. 
In the absence of disorder, the Luttinger-liquid formalism~\cite{carpentier_peca_balents} then yields~$I=\sum_{n=-\infty}^\infty |A_n|^2 I_n$ with
\begin{align}
I_n &=\frac{et_0^2}{\pi^2}\int\mathrm{d}x \int_{-\infty}^0\mathrm{d}t
\ \mathrm{Im}\Big[
\frac{\sin(eVt-(Q-n\kappa)x)}{x-u_0 t - \mathrm{i}\alpha}\  C_n(x,t) 
\Big]\ ,
\label{eqn21}
\end{align}
where
\begin{align}
  C_n(x,t) 
&= \frac{\alpha^{\mu^+_n+\mu^-_n-1}}{(x-u_+ t - \mathrm{i}\alpha)^{\mu^+_n}(x-u_- t + \mathrm{i}\alpha)^{\mu^-_n}}
\label{eqn22a}
\end{align}
with valley-specific exponents~$\mu^\pm_n$ is the zero-temperature correlation function for valley~$n$.
For valley~$n$, e.g., $\mu^+_0=\eta-1$ and $\mu^-_0=\eta$.
Letting $x=-tu$ \cite{carpentier_peca_balents}, we readily integrate over time~$t$. Studying the remaining integral over~$u$,
we identify regions of non-zero current and derive asymptotes for~$eV$ close to $u_\zeta Q$ for $\zeta\in\{\pm,0\}$. 
Characteristic exponents resulting from such calculations (see Appendix A)
are presented in the $I$-$V$ diagrams of Figs.~\ref{fig03}(b) and~(c) for the valleys~$n=0$ and~$1$. 
Numerical evaluation of Eq.~(\ref{eqn21}) leads to the plot in Fig.~\ref{fig03}(a) showing the first four valleys.
The slopes of the non-analytic lines that limit the regions of non-zero current are determined 
by the velocities~$u_\pm$ of the effective modes [Eq.~(\ref{eqn12})].

In the universal limit $\eta=3/2$, the exponents in~(\ref{eqn22a}) for the valley at~$n\kappa$
are given by the simple formulas~$\mu^+_n=(2n+1)^2/2$ and~$\mu^-_n=3/2$ \cite{zuelicke}. The exponent~$\mu^+_n$ of the non-analytic line defined by~$u_-$ 
is smallest for $n=0$ and $-1$. These
are the most relevant terms in~$\hat{\psi}_{2/3}$ [Eq.~(\ref{eqn16})]. 
However, the exponent $\mu^-_n$, which describes the non-analytic behavior at the line defined by~$u_+$, is $n$-independent.
This in the context of Luttinger liquid theory \cite{giamarchi} unusual feature of the spectral function is readily understood: 
Indeed, the number of excited charge modes of a summand in~$\hat{\psi}_{2/3}$ does not depend on~$n$,
which counts neutral modes~$\phi_+\propto \phi_1+3\phi_2$ only.

An exponent~$\mu^-_n=3/2$, universal for all valleys, is reminiscent of the scaling $I\propto V^{1/\nu}$ seen \cite{chang} in tunneling
from a lead into fractional quantum Hall layers regardless of $\nu$ not being a primary filling factor --- an observation contrary to
earlier theoretical predictions \cite{wen_9091}. The point of view of momentum-conserving tunneling
may thus open new possibilities in understanding the observation. 
In the experiment suggested here, the universality of~$\mu^-_n$ may be smeared for large~$n$ due to the increasing exponent at the nearby other non-analytic line 
and, in the presence of disorder, by scattering, whose effect on~$I$ grows with $n$.

We note that other edge models \cite{beenakker,wangmeirgefen} would lead to patterns that differ from the one in Fig. 4. Specifically, the slopes of lines separating the bright (high-current) and dark regions as well as distances between bright regions are model-dependent; the same is true for the exponents~$\mu_n$. The suggested experiment would thus distinguish between possible models.

\paragraph*{Role of disorder.} For strong Coulomb interaction, backscattering off impurities [Eq.~(\ref{eqn13})] is
a relevant perturbation in the renormalization group sense \cite{giamarchi_schulz} and drives the $\nu=2/3$ quantum Hall layer to universal conductance~$G=2e^2/3h$
at low energies \cite{kfp}. In the following discussion, we already assume the low-energy fixed point
and treat $l=2u_+^2/w$ as the effective mean free path \cite{rosenow_halperin}.

\begin{figure}[t]
\centerline{\includegraphics[width=0.8\linewidth]{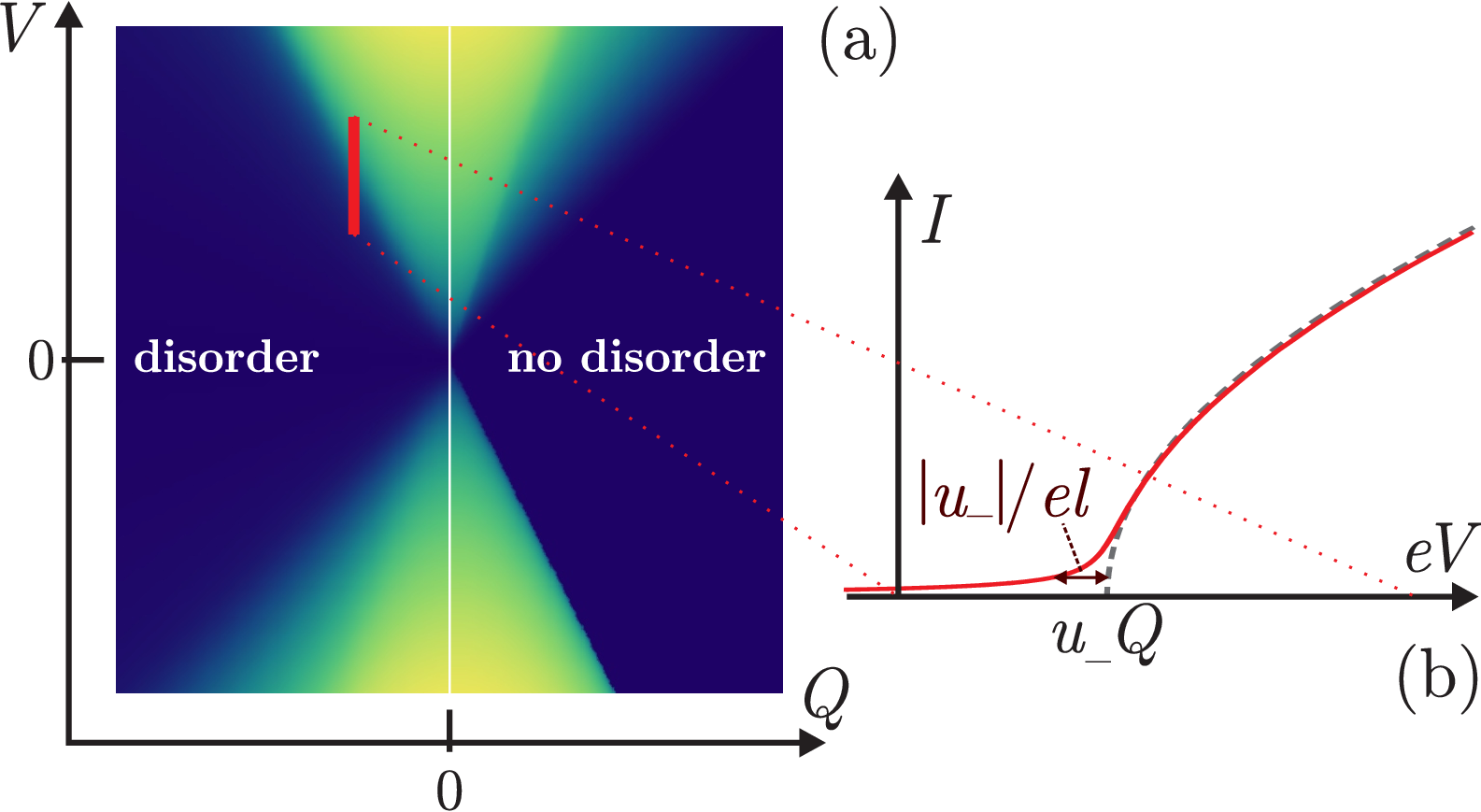}}
\caption{(Color online) (a) Current~$I(V,Q)$ in the clean limit and for weak disorder. (b) $I(V)$ 
for $V$ close to $|u_-Q|$ in the presence and absence (dashed) of disorder.}
\label{fig04}
\end{figure}

Let us have a closer look at voltages $V = u_-Q/e + \delta V$ with small $\delta V$ in the valley~$n=0$. Here the smallest exponent occurs 
($\mu^+_0=1/2$).
As long as disorder is weak compared to the boosting field, $l^{-1} \ll Q$, its effect is merely to blur the non-analytic lines
in the current function~$I(V,B_y)$, see Fig.~\ref{fig04}(a). In fact,
since $Q\sim 10^{7}\ \mathrm{m}^{-1}[B_y/1\ \mathrm{T}]$ and since
experimental evidence (e.g. \cite{bid_mahalu_2009}) 
indicates that $l$ should be at least~$\sim 1\ \mu\mathrm{m}$, the assumption of weak disorder seems valid.
Technically, disorder introduces in the correlation function
$C_0(x,t)$, Eq.~(\ref{eqn22a}), a factor $\exp(-|x|/l)$ (see Appendix B). Close to the non-analytic line~$eV=u_\zeta Q$, however, an explicit evaluation of~$I(V)$
is still possible (see Appendix A). Along the vertical bar in Fig.~\ref{fig04}(a),
\begin{align}
I(\delta V)\simeq \frac{4et_0^2\alpha|u_-|^{\frac{1}{2}}}
{\pi(u_0-u_-)(u_+-u_-)^{\frac{1}{2}}}
\frac{|Q|^{\frac{1}{2}}}{l^{\frac{1}{2}}}\ 
g\Big(
\frac{\delta V}{|u_-|/e l }
\Big)\ .
\label{eqn51}
\end{align}
The shape of the blurring [see Fig.~\ref{fig04}(b)] is described by the function $g(x)=-\mathrm{Im}[(-x-\mathrm{i})^{1/2}]$. 
For $x\gg 1$, $g(x)\simeq \sqrt{x}$, restoring the $I \propto \delta V^{1/2}$ power law of the clean edge at
large bias voltage.
For~$x\rightarrow -\infty$, $g(x)\simeq (-x)^{-1/2}/2$ and we find a power law decay $I \propto l^{-1}(-\delta V)^{-1/2}$ at
bias voltages below~$|u_-Q|$.
In the clean limit, $l\rightarrow\infty$, we recover $I=0$ for $\delta V<0$. 
Finally, as shown in Fig.~\ref{fig04}(b), the function~$I(V)$ in the presence of disorder is slightly displaced in the positive $V$-direction.
This shift $\sim(|u_-|/e)(l^3|Q|)^{-1/2}$ is a next-to-leading order correction to formula~(\ref{eqn51}).

\paragraph*{Conclusion.}
We have suggested a bilayer experimental setup to study the spectral function
of the $\nu=2/3$ state in a two-terminal measurement of current as a function of bias voltage and 
transverse magnetic field. The edge model of two counter-propagating channels may be tested 
by its predicted pattern of current \emph{valleys}. 
We studied the non-analyticities in the current-voltage characteristics 
associated with the channels, finding universal exponents.
Discussing the role of disorder, we quantified its blurring effect for the non-analytical features of the
current found here.
\paragraph*{Acknowledgments.} We thank I. Petkovi\'c for discussions. 
This work was supported by the U.S.-Israel Binational Science
Foundation (Grant 2010366), by NSF DMR-1206612, by the German-Israeli Foundation (GIF), and by the DFG. H.M. acknowledges the Yale Prize Postdoctoral Fellowship.

\newpage

\section{Appendix A: Calculation of tunnel current}

In this Appendix, we present the explicit evaluation of the tunnel current in the absence and presence of disorder. 
We consider here only the~$n=0$ term in the field operator~$\hat{\psi}_{2/3}$ [Eq.~(5)], corresponding to tunneling between the 
probing edge and the $\nu=1$ channel of the fractional edge without exciting additional neutral modes. We are going to focus on bias voltages 
$V = (u_-+\delta u)Q/e$ (where $Q=eB_yd$) mostly in the limit of small~$\delta u$, i.e. $|\delta u| \ll |u_-|,u_+$. We remind the reader that~$u_-<0$.

\subsection{Free chiral electrons}

As a warm-up exercise, we consider tunneling between the probing edge and the~$\nu=1$ channel of the $\nu=2/3$ edge in the absence of both disorder and Coulomb interactions, $u_{12}=0$.
Then, clearly, $u_+=u_2$ and $u_-=-u_1$. In the limit of non-interacting free electrons, we straightforwardly obtain the current per unit length~$I$ using
Fermi's golden rule,
\begin{widetext}
\begin{align}
I &= \frac{2\pi e}{L} \sum_{k,k'} |t_{kk'}|^2 \delta(-u_1(k-Q)-u_0 k + eV)\big\{
f(-u_1(k-Q))\big[1-f(u_0 k)\big]
-
\big[1-f(-u_1(k-Q))\big]f(u_0 k)
\big\}\ ,
\label{fgr01}
\end{align}
\end{widetext}
where~$L$ is the length of the tunnel-coupled edges, $t_{kk'}$ the tunneling matrix element, and $f(\varepsilon)=[\exp(\varepsilon/T)+1]^{-1}$
the Fermi distribution function. For momentum-resolved tunneling, $t_{kk'}=t_0\delta_{kk'}$, cf. Eq.~(4). In the zero-temperature limit $T\rightarrow 0$,
the evaluation of the then trivial integrals in Eq.~(\ref{fgr01}) yields
\begin{align}
I &= \frac{et_0^2}{u_0+u_1}
\big[
\Theta(eV-u_0 Q)-\Theta(-eV -u_1 Q)
\big]
\label{fgr02}\ .
\end{align}
In this formula, $\Theta$ denotes the Heaviside step function. Plotting~$I$, Eq.~(\ref{fgr02}), as a function of~$V$ and~$B_y$
leads to the two dark-colored regions in Fig.~3(a) with constant non-zero current~$I=et_0^2/(u_0+u_1)$ for $V>0$ and $I=-et_0^2/(u_0+u_1)$
for $V<0$.

\subsection{Strongly-correlated edge}

\begin{figure*}[t]
\centerline{\includegraphics[width=0.8\linewidth]{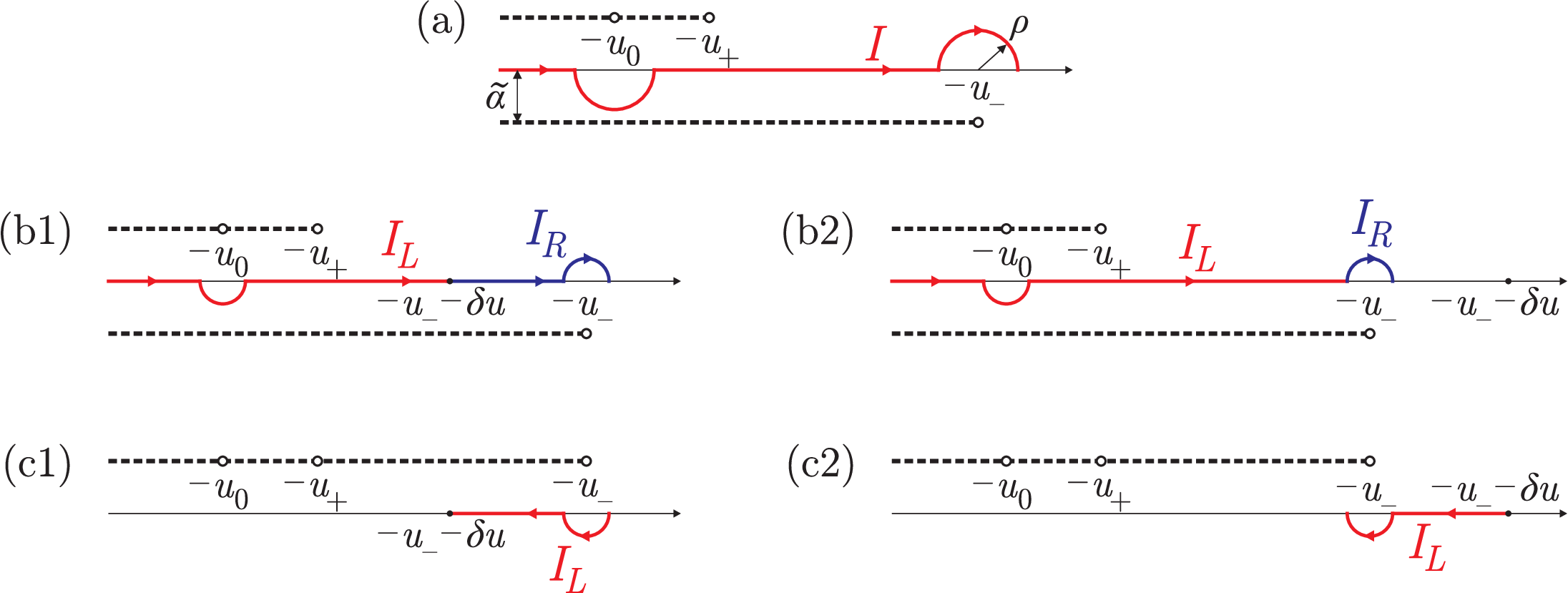}}
\caption{(a) Contour of integration in Eq.~(\ref{calc03}). For~$u>-u_-$, the imaginary part and, hence, the integrand vanishes.
For $\delta u>0$, (b1) shows the decomposition of the integration contour into the two sub-contours for integrals~$I_L$ and~$I_R$ along which the argument of $\mathrm{Im}$ is analytic; (c1) shows the deformed integration contour of the ``reorganized'' integrand
of Eq.~(\ref{calc04}). One finds that $I_L$ and $I_R$ cancel each other. For $\delta V<0$, (b2) shows a convenient decomposition
of the integral~$I$, whose left part~$I_L$ is similarly ``reorganized''. The contour of~$I_L$ is then deformed as shown in (c2). $I_R$ and $I_L$ together produce the finite contribution to the current given in Eq.~(\ref{calc05}).}
\label{fig06}
\end{figure*}

Let us now turn to the physically more realistic model of the strongly-correlated $\nu=2/3$ edge as studied in Ref.~\cite{kfp}.
For simplicity, we assume transverse magnetic fields~$B_y$ such that~$Q>0$. 
We thus start with the following expression for the current~$I=I_0$, cf. Eq.~(6):
\begin{widetext}
\begin{align}
I &= \frac{t_0^2\alpha^{2\eta-2}}{\pi^2}\int_{-\infty}^{0}\mathrm{d}t\int\mathrm{d}x
\sin(eVt-Qx)\exp(-|x|/l)\ \mathrm{Im}\Big[
\frac{1}
{(x-u_0t-\mathrm{i}\alpha)(x-u_+t-\mathrm{i}\alpha)^{\eta-1}(x-u_-t+\mathrm{i}\alpha)^{\eta}}
\Big]\ .
\label{calc01}
\end{align}
The factor~$\exp(-|x|/l)$ is due to disorder. We present a derivation of it starting from the microscopic model
used in the main text in Appendix~B. In the clean limit, $l\rightarrow\infty$
and this factor becomes unity. Transforming the spatial variable as~$x=-tu$ and $\mathrm{d}x=-t\mathrm{d}u$ \cite{carpentier_peca_balents}, we are in the position to immediately perform the integration over time~$t$. We obtain
\begin{align}
I &= \frac{t_0^2\alpha^{2\eta-2}Q^{2\eta-2}}{\pi^2}\int\mathrm{d}u
\ \Gamma(2-2\eta)\ \mathrm{Im}\Big[
\frac{\mathrm{i}}{2}\frac{\mathrm{e}^{\mathrm{i}\pi\eta}(u_-+\delta u + u - \mathrm{i}\varepsilon|u|)^{2\eta-2}-\mathrm{e}^{-\mathrm{i}\pi\eta}(u_-+\delta u + u + \mathrm{i}\varepsilon|u|)^{2\eta-2}}
{(u+u_0-\mathrm{i}\tilde{\alpha})(u+u_+-\mathrm{i}\tilde{\alpha})^{\eta-1}(u+u_-+\mathrm{i}\tilde{\alpha})^{\eta}}
\Big]
\label{calc02}
\end{align}
with~$\varepsilon=1/(lQ)$ and~$\tilde{\alpha}\rightarrow 0^+$.

\subsection{Clean limit}

In the clean limit, $l\rightarrow\infty$ (or~$\varepsilon\rightarrow 0$) and Eq.~(\ref{calc02}) reduces to
\begin{align}
I &= -\frac{t_0^2\alpha^{2\eta-2}Q^{2\eta-2}}{\pi^2}\sin(\pi\eta)\int\mathrm{d}u
\ \Gamma(2-2\eta)\ \mathrm{Im}\Big[
\frac{\sgn(u_-+\delta u + u)|u_-+\delta u + u|^{2\eta-2}}
{(u+u_0-\mathrm{i}\tilde{\alpha})(u+u_+-\mathrm{i}\tilde{\alpha})^{\eta-1}(u+u_-+\mathrm{i}\tilde{\alpha})^{\eta}}
\Big]\ .
\label{calc03}
\end{align}
\end{widetext}
The integration contour of the integral~(\ref{calc03}) lies between two branch cuts, see Fig.~\ref{fig06}(a). Poles of order~$\geq 1$
are at $-u_0+\mathrm{i}\tilde{\alpha}$ and $-u_--\mathrm{i}\tilde{\alpha}$, another pole of order~$<1$ is found at~$-u_++\mathrm{i}\tilde{\alpha}$. 
The numerator introduces a non-analyticity at $u=-u_--\delta u$. We evaluate the integral separately for the two cases~$\delta u>0$ and~$\delta u<0$. In the following, we interpret for each step the order of integration over~$u$ and taking the imaginary part
in the way as it is more convenient, since these two operations commute.

\paragraph*{Case $\delta u >0$.}
In the case~$\delta u>0$, we cut the integration contour at the non-analytic point $u=-u_--\delta u < -u_-$ into a left ($L$) and right ($R$) part,
see Fig.~\ref{fig06}(b1).
For the left part, we ``reorganize'' the integrand as
\begin{widetext}
\begin{align}
I_L &= \frac{t_0^2\alpha^{2\eta-2}Q^{2\eta-2}}{\pi^2}\sin(\pi\eta)\int_{-\infty}^{-u_--\delta u}\mathrm{d}u
\ \Gamma(2-2\eta)\ \mathrm{Im}\Big[
\frac{(u_-+\delta u + u-\mathrm{i}\tilde{\alpha})^{2\eta-2}}
{(u+u_0-\mathrm{i}\tilde{\alpha})(u+u_+-\mathrm{i}\tilde{\alpha})^{\eta-1}(u+u_--\mathrm{i}\tilde{\alpha})^{\eta}}
\Big]\ .
\label{calc04}
\end{align}
\end{widetext}
In the form of Eq.~(\ref{calc04}), all poles and branch cuts are located in the upper half-plane and we may deform the contour
as shown in Fig.~\ref{fig06}(c1). Investigating~$I_L$ along the new contour, we quickly recognize that $I_L=-I_R$ so that for $\delta u >0$,
we find
\begin{align}
I(\delta u>0)=0\ .
\label{calc04a}
\end{align}
Note that this holds for all $\delta u$ with $0<\delta u < u_+-u_-$, independently from whether~$\delta u$ is small.

\paragraph*{Case $\delta u <0$.}
In the case~$\delta u<0$, the non-analyticity of the numerator lies to the right of~$-u_-$. We thus split the integration contour at $u=-u_--\rho$ with~$\rho\rightarrow 0$ the radius of the semi-circle around the pole at~$-u_-$, see Fig.~\ref{fig06}(b2). For the left part,
we reorganize the poles as in Eq.~(\ref{calc04}) and deform the integration contour as in Fig.~\ref{fig06}(c2). Contrarily to the case $\delta u>0$, there is evidently no cancellation of~$I_L$ and $I_R$. We calculate~$I_R$ and $I_L$ in the limit of small~$\delta u$ separately for finite~$\rho$, leading to two divergent contributions in the limit~$\rho\rightarrow 0$. The sum~$I_L+I_R$, however, is regular as it should be, and reexpressing $\delta u$ in terms of a small (negative) voltage~$\delta V=Q\delta u/e$, we find
\begin{align}
I(\delta V) &\simeq
-\frac{C_{-} e t_0^2\alpha^{2\eta-2}}
{(u_0-u_-)(u_+-u_-)^{\eta-1}}
Q^{\eta-1} (-e\delta V)^{\eta-1}
\label{calc05}
\end{align}
with $C_-=4^{\eta-1}\Gamma(\eta-\tfrac{1}{2})/(\pi^{1/2}\Gamma(\eta)\Gamma(2\eta-1))$. In particular, in the universal limit $\eta=3/2$, we find
\begin{align}
I(\delta V) &\simeq
-\frac{4 e t_0^2\alpha}
{\pi(u_0-u_-)(u_+-u_-)^{\frac{1}{2}}}
Q^{\frac{1}{2}} (-e\delta V)^{\frac{1}{2}}\ .
\label{calc06}
\end{align}
Equations~(\ref{calc04a}) and~(\ref{calc06}) accurately describe the $I$-$V$ characteristic in the asymptotic limit close to voltages~$V= Qu_-/e$, cf. Fig.~4(b).

\subsection{Finite mean free path}

In case of a finite mean free path~$l$, the numerator in Eq.~(\ref{calc02}) introduces two more branch cuts. As a result, an analytical evaluation for a general~$\eta$ becomes very difficult even in the asymptotic limit. In
the universal limit~$\eta=3/2$, the situation is simpler and allows for a rather straightforward asymptotic calculation.
Furthermore, we are assuming a large mean free path $l\gg Q^{-1}$, i.e. $\varepsilon=1/(lQ)\ll 1$.
Taking the limit~$\eta \rightarrow 3/2$ from below in the integrand of Eq.~(\ref{calc02}), we find
\begin{widetext}
\begin{align}
I &= -\frac{t_0^2\alpha Q}{2\pi^2}\int\mathrm{d}u
\ \mathrm{Im}\Big[
\frac{W(u)}
{(u+u_0-\mathrm{i}\tilde{\alpha})(u+u_+-\mathrm{i}\tilde{\alpha})^{\frac{1}{2}}(u+u_-+\mathrm{i}\tilde{\alpha})^{\frac{3}{2}}}
\Big]
\label{calc11}
\end{align}
with
$W(u)=W_1(u)+W_2(u)+W_{\varepsilon=0}(u)$
where
\begin{align}
W_1(u) &= \pi\varepsilon|u|\sgn(u+u_-+\delta u)\ ,
\label{calc13a}
\\
W_2(u) &= 
 (u+u_-+\delta u)\ln\Big[
 \frac{(u+u_-+\delta u)^2+\varepsilon^2u^2}{(u+u_-+\delta u)^2}
 \Big]
 -\mathrm{i}\varepsilon|u|\ln\Big[
 \frac{u+u_-+\delta u-\mathrm{i}\varepsilon|u|}{u+u_-+\delta u+\mathrm{i}\varepsilon|u|}
 \Big]\ ,
\label{calc13b}
\\
W_{\varepsilon=0}(u) &= (u+u_-+\delta u)
 \ln\Big[
 \frac{(u+u_-+\delta u)^2}{(u+u_+-\mathrm{i}\tilde{\alpha})(u+u_-+\mathrm{i}\tilde{\alpha})}
 \Big]\ .\label{calc13c}
\end{align}
\end{widetext}
Accordingly, we split~$I=J_1+J_2+J_{\varepsilon=0}$. The integral of~$J_1$ can be calculated using the
same strategies as in the clean limit,
\begin{align}
J_1 &= - \frac{2t_0^2\alpha Q}{\pi(u_0-u_-)(u_+-u_-)^{\frac{1}{2}}} 
 \ u_-\varepsilon\ \frac{\Theta(\delta u)}{|\delta u|^{\frac{1}{2}}}
 + \Delta I
\label{calc26}
\end{align}
with
\begin{align}
\Delta I &= \frac{t_0^2\alpha}{2\pi}\ \varepsilon\ \Big[
\frac{\pi u_0}{(u_0-u_+)^{\frac{1}{2}}(u_0-u_-)^{\frac{3}{2}}}\label{calc27}\\
&\quad -\int_0^{u_+}\frac{\mathrm{d}u\ (u_+-u_-)}{u^{\frac{1}{2}}[u+(u_0-u_+)][-u+(u_+-u_-)]^{\frac{3}{2}}}
\Big]\ .
\nonumber
\end{align}
Equation~(\ref{calc26}) constitutes the leading correction to the clean result of Eqs.~(\ref{calc04a}) and~(\ref{calc06}) 
in the intermediate regime of $\varepsilon|u_-|\ll |\delta u| \ll |u_-|$ but is completely inaccurate
for $\delta u\rightarrow 0$ where it diverges as $\Theta(\delta u)|\delta u|^{-1/2}$. Here, $J_2$ has to be taken into account. $J_2$ itself 
is again difficult to evaluate because of the various branch cuts due to $W_2(u)$. It is possible, though, to 
avoid these difficulties by evaluating $\partial^2J_2/\partial\varepsilon^2$ instead and recover
the current~$I$ as
\begin{align}
I = J_1 + \int_0^\varepsilon\mathrm{d}\varepsilon'\int_0^{\varepsilon'}\mathrm{d}\varepsilon''
\ \frac{\partial^2J_2}{\partial\varepsilon^2}\Big|_{\varepsilon=\varepsilon''}\ .
\label{calc21}
\end{align}
The lower limits of~$0$ in the integrations actually become clear only during the subsequent analysis. They are imposed by the necessity
to compensate for the~$|\delta u|^{-1/2}$ divergency in~$J_1$ and the requirement to reproduce
Eqs.~(\ref{calc04a}) and~(\ref{calc06}) in the limit~$\varepsilon\rightarrow 0$.

Using
\begin{align}
\frac{\partial^2 W_2(u)}{\partial\varepsilon^2} &= -\frac{2u^2(u+u_-+\delta u)}{(u+u_-+\delta u)^2+\varepsilon^2u^2}\ ,
\label{calc22}
\end{align}
and $\mathrm{Im}[(u+u_-+\mathrm{i}0^+)^{-\frac{3}{2}}]=\mathrm{Im}[\mathrm{i}(-u-u_-+\mathrm{i}0^+)^{-\frac{3}{2}}]$,
we find that
\begin{widetext}
\begin{align}
\frac{\partial^2J_2}{\partial\varepsilon^2} &= -\frac{t_0^2\alpha Q}{\pi^2}\int\mathrm{d}u
\ \mathrm{Im}\Big[\frac{u^2(u+u_-+\delta u)}{(u+u_-+\delta u)^2+\varepsilon^2u^2}\ 
\frac{\mathrm{i}}
{(u+u_0-\mathrm{i}\tilde{\alpha})(u+u_+-\mathrm{i}\tilde{\alpha})^{\frac{1}{2}}(-u-u_-+\mathrm{i}\tilde{\alpha})^{\frac{3}{2}}}
\Big]
\label{calc23}
\end{align}
no longer contains branch cuts in the lower half-plane. Instead, the lower half-plane now only features a single pole
at $-(u_-+\delta u)(1-\mathrm{i}\varepsilon)/(1+\varepsilon^2)$. Closing the contour around this pole, we 
obtain in the leading order in both~$\varepsilon$ and~$\delta u$ the expression
\begin{align}
\frac{\partial^2J_2}{\partial\varepsilon^2} &\simeq -\frac{t_0^2\alpha Q}{\pi}\ 
\frac{u_-^2}{(u_0-u_-)(u_+-u_-)^{\frac{1}{2}}} 
\mathrm{Im}\Big[
-\frac{1}{(\delta u - \mathrm{i}\varepsilon u_-)^{\frac{3}{2}}}
\Big]\ .
\label{calc24}
\end{align}
Integrating twice over~$\varepsilon$ as in Eq.~(\ref{calc21}), we find
\begin{align}
\int_0^\varepsilon\mathrm{d}\varepsilon'\int_0^{\varepsilon'}\mathrm{d}\varepsilon''
\ \frac{\partial^2J_2}{\partial\varepsilon^2}\Big|_{\varepsilon=\varepsilon''}
&= 
\frac{4t_0^2\alpha Q}{\pi(u_0-u_-)(u_+-u_-)^{\frac{1}{2}}} 
\Big\{
\mathrm{Im}
\big[
(\delta u - \mathrm{i}\varepsilon u_-)^{\frac{1}{2}}
\big]
+ \frac{u_-\varepsilon}{2}\ \frac{\Theta(\delta u)}{|\delta u|^{\frac{1}{2}}}
\Big\}
\ .
\label{calc25}
\end{align}
\end{widetext}
The last term just cancels the divergency coming from~$J_1$, Eq.~(\ref{calc26}). Since the first term is a smooth function of the real variable~$\delta u$, we find
that as a result the entire expression for the current, given by Eq.~(\ref{calc21}), is regular.

Thus combining Eqs.~(\ref{calc26}), (\ref{calc21}), and (\ref{calc25}), the leading-order current at bias voltage~$V=u_-Q/e+\delta V$ for small~$\delta V$, positive~$Q=eB_yd$, and weak disorder ($l\gg Q^{-1}$) is given by
\begin{align}
I(\delta V)
&\simeq 
\frac{4t_0^2\alpha|u_-|\ (Q/l)^{\frac{1}{2}}}
     {\pi(u_0-u_-)(u_+-u_-)^{\frac{1}{2}}} 
\mathrm{Im}
\Big[
\Big(\frac{\delta V}{|u_-|/el} + \mathrm{i}\Big)^{\frac{1}{2}}
\Big]
\ .
\label{calc28}
\end{align}
The constant $\Delta I$, Eq.~(\ref{calc27}), is a next-to-leading order correction in the asymptotic limit considered. It notably 
leads to a horizontal shift of the current as a function of voltage as seen in Fig.~5 for larger~$\delta V$.

For $Q<0$, we obtain analogously the formula displayed in Eq.~(8) in the main text. Clearly, for~$l\rightarrow\infty$, we recover from Eq.~(\ref{calc28}) the result for the clean limit given by Eqs.~(\ref{calc04a}) and~(\ref{calc06}).

\section{Appendix B: Disorder averaging of the universal conductance}

In this Appendix, we derive the prefactor of~$\exp(-|x|/l)$ that decorates the correlation function~$C_n(x,t)$, Eq.~(6),
for~$n=0$ and thus the integrand of Eq.~(\ref{calc01}) in the presence
of disorder. We assume the limit of universal conductance~$G=2e^2/3h$, i.e. $\eta=3/2$. The function~$C_0(x,t)$ is defined as
\begin{align}
C_0(x,t) &= \frac{1}{\mathrm{i}\alpha}
 \big\langle\big\langle
   \exp\big\{\mathrm{i}[\phi_1(0,0)-\phi_1(x,t)]\big\}
 \big\rangle\big\rangle_{\mathrm{dis}}
\label{dis01}
\end{align}
where~$\langle\langle\ldots\rangle\rangle_{\mathrm{dis}}$ denotes, starting from the inside, quantum averaging and then ensemble averaging over disorder.

Using 
the decomposition of~$\phi_1$ into the charge and neutral modes~$\phi_-$ and~$\phi_+$, cf. Eq.~(2), we obtain
\begin{align}
C_0(x,t) = \frac{1}{\mathrm{i}\alpha}
 &\big\langle
   \mathrm{e}^{-\mathrm{i}\sqrt{3/2}\ \phi_-(0,0)}
   \mathrm{e}^{\mathrm{i}\sqrt{3/2}\ \phi_-(x,t)}   
 \big\rangle
\nonumber\\ 
 \times&\big\langle\big\langle
   \mathrm{e}^{\mathrm{i}\phi_+(0,0)/\sqrt{2}}
   \mathrm{e}^{-\mathrm{i}\phi_+(x,t)/\sqrt{2}}   
 \big\rangle\big\rangle_{\mathrm{dis}} 
\label{dis02}\ .
\end{align}
Only~$\phi_+$ is affected by disorder, cf. Eq.~(3). 

In order to perform the disorder-averaging,
we make use of the hidden $\mathrm{SU}(2)$-symmetry pointed out by Kane, Fisher, and Polchinski (KFP)
in Ref.~\cite{kfp} and refermionize the neutral sector into effective pseudo-spin-$\tfrac{1}{2}$ quasiparticles~$\Psi$.
KFP made this refermionization transparent by introducing a mode of bosonic ghosts~$\chi$
identical to~$\phi_+$ except for the fact that~$\chi$ is not affected by the disorder potential.
The ``refermionization identities'' then read~$\Psi_\uparrow=\exp[\mathrm{i}(\chi+\phi_+)/\sqrt{2}]$
and $\Psi_\downarrow=\exp[\mathrm{i}(\chi-\phi_+)/\sqrt{2}]$. Then
\begin{align}
 \big\langle\big\langle
   \mathrm{e}^{\mathrm{i}\phi_+(0,0)/\sqrt{2}}
   \mathrm{e}^{-\mathrm{i}\phi_+(x,t)/\sqrt{2}}   
 \big\rangle\big\rangle_{\mathrm{dis}}
&= \frac{
 \big\langle\big\langle
   \Psi^\dagger_\downarrow(0,0)
   \Psi_\downarrow(x,t)  
 \big\rangle\big\rangle_{\mathrm{dis}}
     }
     {
 \big\langle
   \mathrm{e}^{-\mathrm{i}\chi(0,0)/\sqrt{2}}
   \mathrm{e}^{\mathrm{i}\chi(x,t)/\sqrt{2}}   
 \big\rangle
     }
\label{dis03}
\end{align}
where the correlation function involving~$\boldsymbol{\Psi}=(\Psi_\uparrow,\Psi_\downarrow)$
is obtained from quantum-averaging with respect to the Hamiltonian
\begin{align}
 \hat{H}_{\Psi}
&= \int\mathrm{d}x\ \boldsymbol{\Psi}^\dagger
  \Big[
  -\mathrm{i}u_+\nabla
  + \Xi
  \Big]
  \boldsymbol{\Psi}
\label{dis04}
\end{align}
with
\begin{align}
\Xi(x) &= \left(\begin{array}{cc}
  0 & \xi^*(x) \\ 
  \xi(x) & 0
  \end{array}\right)
\label{dis04a}  \ .
\end{align}
For the ensemble average, we assume Gaussian disorder with
the correlation function
\begin{align}
\big\langle\xi(x)\xi^*(x')\big\rangle_\mathrm{dis}= w\delta(x-x')
\label{dis04b}\ .
\end{align}
By the unitary disorder-dependent gauge transformation
\begin{align}
\Psi(x) &= \tilde{U}(x)\tilde{\Psi}(x)
\label{dis05}
\end{align}
with
\begin{align}
\tilde{U}(x) = \mathrm{T}_x\exp\Big[-
\mathrm{i}u_+^{-1}\int^x_0\mathrm{d}x'\ \Xi(x')
\Big]
\label{dis06}
\end{align}
we obtain a diagonal Hamiltonian without randomness,
\begin{align}
 \hat{H}_{\tilde{\Psi}}
&= \int\mathrm{d}x\ \tilde{\boldsymbol{\Psi}}^\dagger
     (-\mathrm{i}u_+\nabla)
  \tilde{\boldsymbol{\Psi}}
\label{dis07}\ ,
\end{align}
for the fermions~$\tilde{\Psi}$. 

The disorder, which the physical neutral mode, represented by~$\Psi$, experiences,
thus drops out of the quantum average,
\begin{align}
&\big\langle\big\langle
   \Psi^\dagger_\downarrow(0,0)
   \Psi_\downarrow(x,t)  
 \big\rangle\big\rangle_{\mathrm{dis}}\nonumber\\
&\quad=\big\langle
   \tilde{\Psi}^\dagger_\downarrow(0,0)
   \tilde{\Psi}_\downarrow(x,t)  
 \big\rangle
 \mathrm{tr}
 \Big[
 \left(\begin{array}{cc}
  0 & 0 \\ 
  0 & 1
  \end{array}\right)
  \big\langle\tilde{U}(x)\big\rangle_{\mathrm{dis}}
 \Big]
\label{dis08}\ .
\end{align}
We average the $x$-ordered exponential~$\tilde{U}(x)$, Eq.~(\ref{dis06}),
introducing a proper discretization,
\begin{align}
\tilde{U}(x) = \lim_{N\rightarrow\infty} \prod_{j=0}^{N-1}\exp\big(
-\mathrm{i}u_+^{-1}\Delta \Xi_j
\big)
\label{dis09}
\end{align}
where $\Delta=|x|/N$ and $\Xi_j = \sgn(x)\Xi(j\Delta-|x|\Theta(-x))$. Each exponential function
is readily evaluated,
\begin{align}
&\exp\big(
-\mathrm{i}u_+^{-1}\Delta \Xi_j
\big)\label{dis10}\\
&\quad =
\left(\begin{array}{cc}
  \cos(u_+^{-1}\Delta|\xi_j|) & -\xi_j^*\sin(u_+^{-1}\Delta|\xi_j|)/|\xi_j| \\ 
  -\xi_j\sin(u_+^{-1}\Delta|\xi_j|)/|\xi_j| & \cos(u_+^{-1}\Delta|\xi_j|)
  \end{array}\right)\ .
\nonumber
\end{align}
The correlation~(\ref{dis04b})
translates as~$\langle\xi_j\xi_{j'}^*\rangle_{\mathrm{dis}}=w\delta_{jj'}/\Delta$.
Averaging Eq.~(\ref{dis10}) over disorder, the matrix structure becomes trivial
and we find
\begin{align}
\big\langle\exp\big(
-\mathrm{i}u_+^{-1}\Delta \Xi_j
\big)\big\rangle_{\mathrm{dis}}
&=
1-u_+^{-1}\sqrt{\Delta w}\ F(u_+^{-1}\sqrt{\Delta w}/2)
\label{dis11}
\end{align}
where
\begin{align}
F(z)=\mathrm{e}^{-z^2}\int_0^z\mathrm{d}\zeta\ \mathrm{e}^{\zeta^2}
 \label{dis12}
\end{align}
is the Dawson integral. For small~$z$, $F(z)\simeq z$ so that
for the average of Eq.~(\ref{dis09}) we obtain
\begin{align}
\big\langle\tilde{U}(x)\big\rangle_{\mathrm{dis}} &= \lim_{N\rightarrow\infty} 
\big(
1- u_+^{-2}w\Delta/2
\big)^N = \exp(-|x|/l)
\label{dis13}
\end{align}
where $l=2u_+^2/w$ is the mean free path. Inserting Eq.~(\ref{dis13}) into Eq.~(\ref{dis08}),
we can go back all steps to Eq.~(\ref{dis01}), replacing everywhere the disorder-average by a multiplication
with the factor~$\exp(-|x|/l)$. As a result,
\begin{align}
C_0(x,t) &= C_0(x,t) \Big|_{l\rightarrow\infty}\ \exp(-|x|/l)
\label{dis14}\ .
\end{align}
We note that, e.g., for the correlation function $C_1(x,t)$ in Eq.~(6),
which describes the propagation in the $\nu=1/3$ edge channel, disorder produces the factor $\exp(-9|x|/l)$. 
Generally, the effective mean free path in the valley at $n\kappa$
is by a factor of $(2n+1)^2$ smaller than in the valley for $n=0$ so that blurring effects due to disorder are enhanced for large~$n$.

\end{document}